\documentclass[prb,twocolumn,showpacs,preprintnumbers,amsmath,amssymb,superscriptaddress,aps]{revtex4-1}

\usepackage{amsmath,amssymb,amsfonts} 	
\usepackage{graphicx}
\usepackage{hhline}
\usepackage[latin1]{inputenc}
\usepackage{subfigure}

\begin{document}

\title{Designing in-plane heterostructures of quantum spin hall insulators from first principles: 1T'-MoS$_2$ with adsorbates.}

\author{Thomas Olsen}
\email{tolsen@fysik.dtu.dk}
\affiliation{Center for Atomic-Scale Materials Design and Center for Nanostructured Graphene (CNG), Department of Physics, Technical University of Denmark}

\begin{abstract}
Interfaces between normal and topological insulators are bound to host metallic states that are protected by time-reversal symmetry and are therefore robust against disorder and interface reconstruction. Two-dimensional topological insulators (quantum spin Hall insulators) offer a unique opportunity to change the local topology by adsorption of atoms or molecules and thus comprise an ideal platform for designing topological heterostructures. Here, we apply first principles calculations to show that the quantum spin Hall insulator 1T'-MoS$_2$ exhibits a phase transition to a trivial insulator upon adsorption of various atoms. It is then demonstrated that one-dimensional metallic boundary states indeed arise in a ribbon geometry of alternating regions with and without adsorbed oxygen and that these boundary states generically constitute simple linear connections between valence and conduction bands. This is in sharp contrast to topological edge states that typically exhibit strong dispersion that are sensitively to a particular edge termination. The heterostructure is also suggestive of a simple design of one-dimensional metallic networks in sheets of 1T'-MoS$_2$.
\end{abstract}
\maketitle

\section{Introduction}
The topological classification of insulators implies that non-trivial physics may arise at the interface of two materials exhibiting different band topology.\cite{Hasan2010} Specifically, if the topology is protected by a certain symmetry and the interface respects that symmetry, the interface will host a metallic state, since this is the only means by which the topology can change across the interface. Regarding vacuum as a special case of a trivial insulator, it follows that metallic states may be hosted at the edges and surfaces of two-dimensional (2D) and three-dimensional (3D) topological insulators respectively. In particular, topological insulators protected by time-reversal symmetry,\cite{Kane2005,Kane2005a,Fu2007a} are guaranteed to host metallic states at any non-magnetic edge or surface; irrespective of the details of surface termination. Several 3D materials have now been demonstrated to exhibit a non-trivial band topology protected by time-reversal symmetry\cite{Bansil2016} (simply referred to as topological insulators in the following) and the associated surface states has been studied theoretically\cite{Teo2008, Zhang2009} as well as experimentally.\cite{Hsieh2009, Hsieh2008, DiPietro2013} The topological surface states are protected from impurity scattering by time-reversal symmetry and topological insulators thus constitute a promising candidate for dissipationless electronics applications.

In a different line of development, the past decade has witnessed a rapidly increasing interest in 2D materials. Starting with graphene,\cite{Novoselov2004} the focus rapidly broadened to include several graphene derivatives as well as transition metal dichalcogenides (TMDs) and hexagonal boron nitride (hBN).\cite{Novoselov2005} These materials has a large number of properties that significantly deviate from there 3D counterparts. For example, the electronic screening in 2D is much less efficient than in 3D\cite{Cudazzo2011} and gives rise to qualitatively different plasmon dispersion for 2D metals\cite{Yan2011a} and large exciton binding energies in 2D semiconductors.\cite{Pulci2012, Huser2013b, Chernikov2015a, Olsen2016} Another interesting property is the coupling between valley indices and angular momentum in the TMDs,\cite{Xiao2012} which allows for optical control of the valley degrees of freedom\cite{Mak2012, Behnia2012, Jones2013} as well as a realization of the valley Hall effect.\cite{Mak2014, Olsen2015} However, the most intriguing property of the 2D materials is the possibility of tuning specific properties such as the band gap. This can be accomplished either by constructing stacks of different 2D layers\cite{Geim2013, Andersen2015, Latini2015} or by simply adsorbing various atoms or molecules on the face of a single layer. In the present work it will be demonstrated that the latter approach can also be applied to change the topology of 2D materials.

Graphene can rightfully be regarded as the parent material for the host of known 2D materials today, but it is interesting to note that graphene also played a prominent role in the theoretical development of topological band theory. Initially Haldane showed that it is possible to obtain a quantum anomalous Hall insulator from a model of graphene with a time-reversal breaking second-nearest neighbor interaction.\cite{Haldane1988} Subsequently, Kane and Mele predicted that spin-orbit interaction opens a topological gap in graphene, which thus comprises the first prediction of a quantum spin Hall insulator (QSHI).\cite{Kane2005a} However, due to the weak spin-orbit interaction in graphene, the gap is too small to be measured and the prediction has not been verified experimentally. Likewise, first principles calculations have shown that silicene, germanene, and stanene are QSHIs. In these systems, the spin-orbit coupling and the topological gap is much larger (25-75 meV) than in the case of graphene, but experimental verification of the quantum spin Hall insulating phase is hindered by the fact that synthesis of the materials requires growth on a substrate, which significantly alters the electronic properties. Recently, Qian et al.\cite{Qian2014} showed that the class of TMDs MX$_2$ (M=Mo,W X=S,Se) in the 1T' structure are all QSHIs with gaps in the range 50-100 meV and subsequently a different family of TMDs, known as haeckelites, were also shown to be QSHIs with gaps on the order of 10-50 meV.\cite{Nie2015} It has recently been shown that sheets of 1T'-MoS$_2$ nanosheets can be chemically exfoliated\cite{Voiry2013, Lin2014, Acerce2015} and this novel 2D material thus comprises a promising candidate for a metastable QSHI that can be studied experimentally.

\begin{figure}[tb]
   \includegraphics[width=7.7 cm]{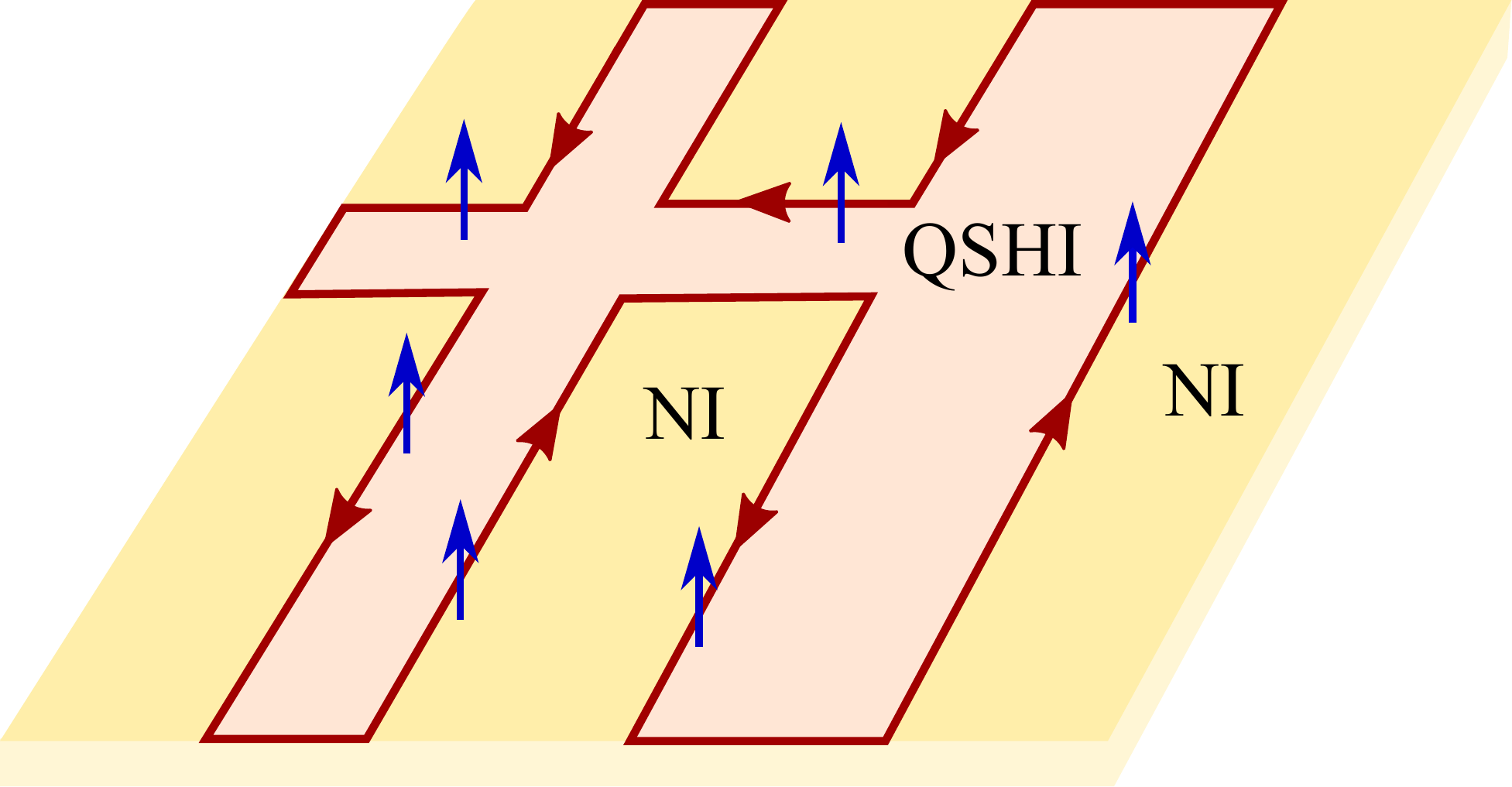}
\caption{(Color online) In-plane heterostructure of a quantum spin Hall insulator (QSHI) and a normal insulator (NI). The boundary regions host one-dimensional spin-polarized metallic states. In addition to the boundary state shown here there will be a counter-propagating state of opposite spin.}
\label{fig:drawing}
\end{figure}

Compared to their 3D topological counterparts, the 2D QSHIs are particularly interesting because they support one-dimensional (1D) metallic edge states. It is well known that any amount of disorder leads to Anderson localization in 1D,\cite{flensberg} which effectively implies that strictly 1D metals cannot be realized in real materials. However, the edges of quantum spin Hall insulators provide a loophole, since the states are protected from impurity scattering by time-reversal symmetry and the QSHI thus constitutes a unique possibility to study conductivity in 1D. For the purpose of gaining optimal control over the edge states it is desirable to have a single pair of metallic states at a particular edge that each cross the Fermi level once.\cite{Hasan2010} This is the minimal requirement by the topology, but the presence of additional non-topological edge state may obscure the picture and the number of such states will generically be highly sensitive to the details of surface termination, which is difficult to control at the atomic level. Moreover, the topological edge states will often originate from dangling bonds and have a strongly dispersive behavior with several Fermi level crossings.\cite{Nie2015} In contrast, the non-topological edge states and multiple Fermi level crossings are likely to be eliminated if one considers a heterostructure of a trivial 2D insulator and a QSHI. In that case there will be no dangling bonds and the topological boundary states will typically exhibit simple linear dispersion. Furthermore, if the design of such heterostructures could be controlled in detail it would be possible to construct electronic circuits like the one shown in Fig. \ref{fig:drawing}, which is solely based one 1D topological boundary states. We note that a similar construction has been proposed, in which a mesh of metallic graphene ribbons are embedded into an insulating sheet of hBN,\cite{Liu2013} but that will only result in quasi-1D metallic channels that may still be subject to Anderson localization.

In the present work, we will apply density functional theory (DFT) to show that the topology of 2D materials can be changed by adsorbing atoms onto known QSHIs. This suggests a simple route to the design of heterostructures; namely by adding adsorbates on local regions in an intrinsic QSHI. In Sec. \ref{sec:results}, we will apply this construction and show that oxygen atoms provides a simple and effective means to change the topology of 1T'-MoS$_2$. We then study the topological boundary states in heterostructures resulting from a local adsorption of oxygen and show that the 1D metallic states are indeed well-behaved and do not exhibit multiple Fermi level crossings. In the appendix we document the implementation of spin-orbit coupling in the electronic structure code GPAW\cite{Enkovaara2010a}, the implementation of an interface to the Wannier90 package\cite{Marzari2012} and summarize the equations used for the iterative Greens function approach used to obtain the spectral function of the heterostructures.

\section{Results}\label{sec:results}
All DFT calculations in the present section were obtained with the electronic structure code GPAW,\cite{Enkovaara2010a} which is based on the projector augmented wave methodology.\cite{blochl} The calculations were performed with the PBE functional\cite{Perdew1996} using a plane wave basis and a 600 eV cutoff. We used an $8\times8$ $k$-point mesh for the simple unit cell of 1T'-MoS$_2$ and corresponding $k$-point densities for calculations of larger structures. Spin-orbit coupling was added as a non-selfconsistent correction to the band structures and eigenstates. We refer to appendix \ref{app:so} for details on the implementation in GPAW.

Surface and boundary spectral functions were obtained using an iterative Greens function scheme.\cite{Sancho2000, Sancho2000a} To accomplish this the Hamiltonian was first transformed to a local basis of Wannier functions using the Wannier90 software package.\cite{Marzari2012} We refer to appendix \ref{app:wannier} for details on the implementation of the GPAW-Wannier90 interface.
\begin{figure}[b]
   \includegraphics[width=6.0 cm]{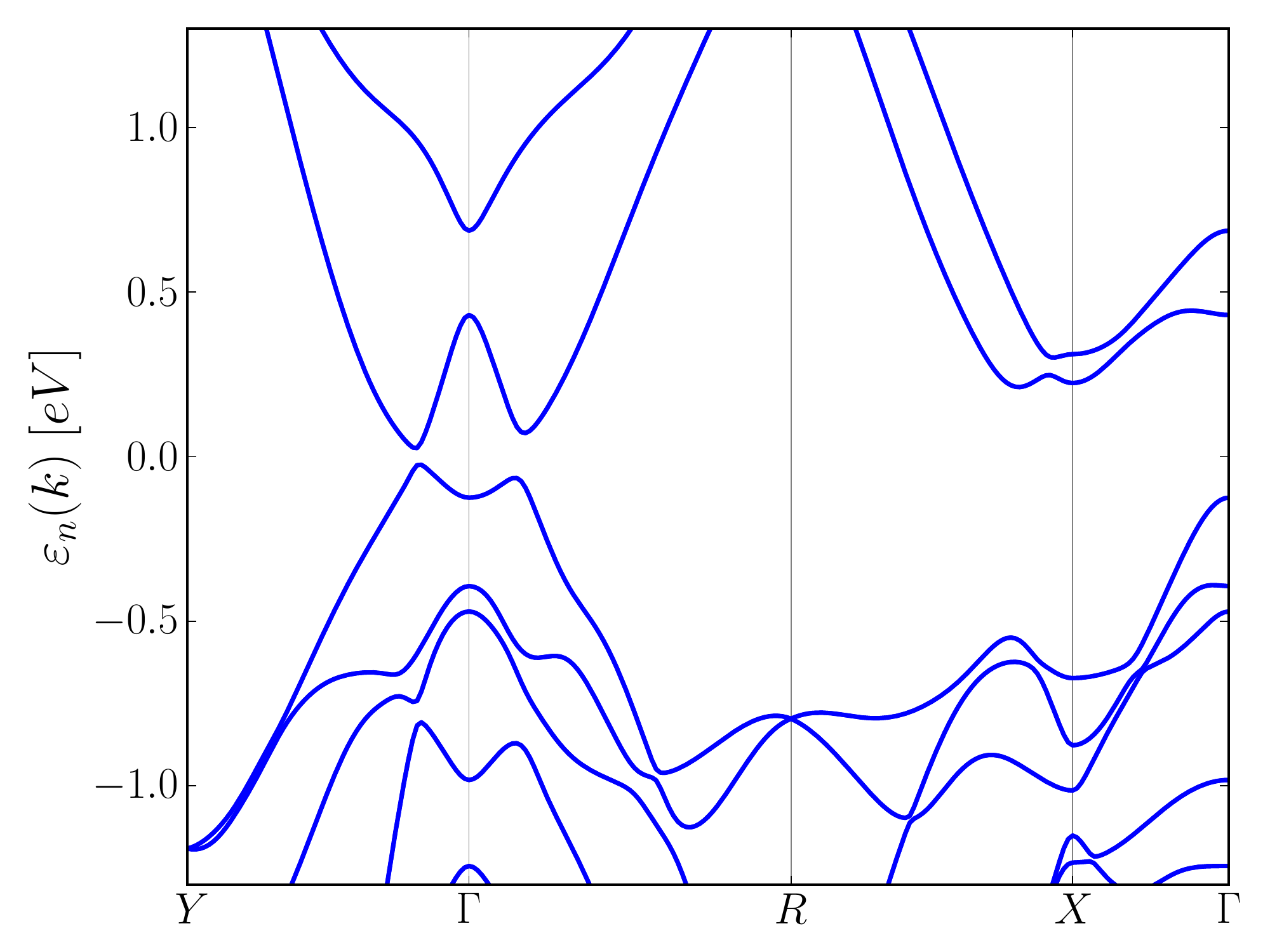}
   \includegraphics[width=2.2 cm]{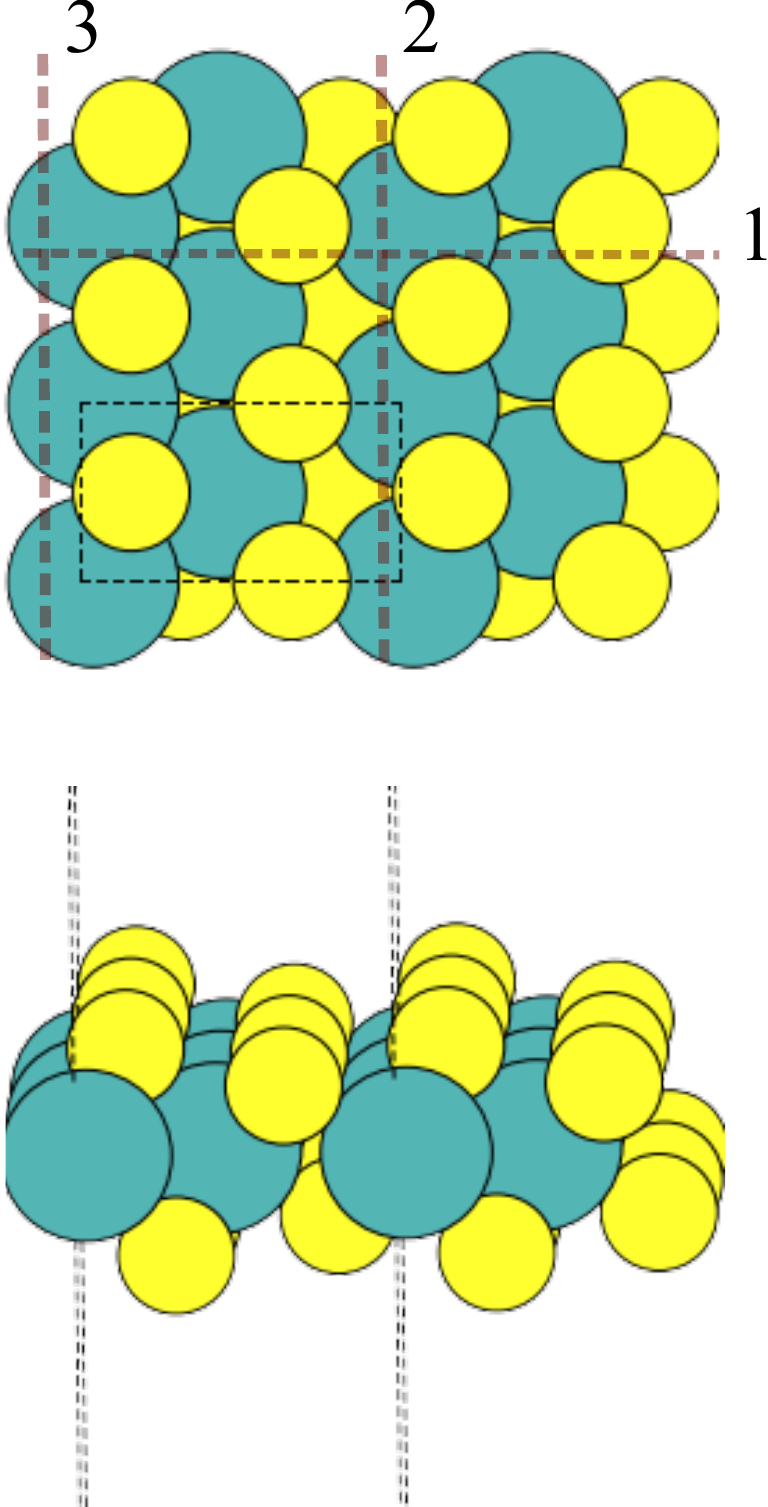}
\caption{(Color online) Left: band structure of pristine 1T'-MoS$_2$. Right: top and side view of the 1T'-MoS$_2$ structure. Unit cell indicated by black dashed lines and different edge terminations indicated by the three cuts.}
\label{fig:bands_1T}
\end{figure}

\begin{figure*}[tb]
   \includegraphics[width=5.8 cm]{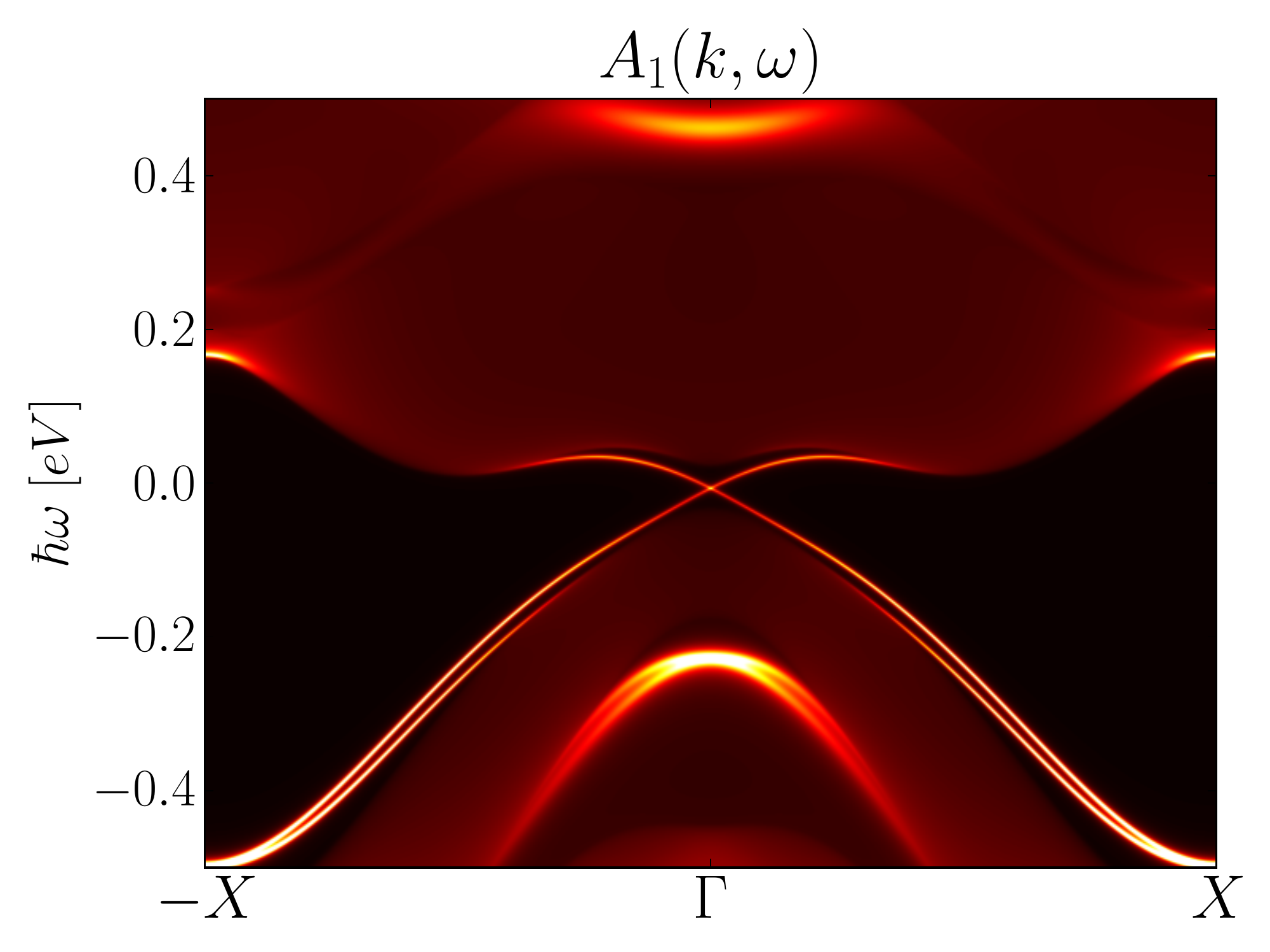}
   \includegraphics[width=5.8 cm]{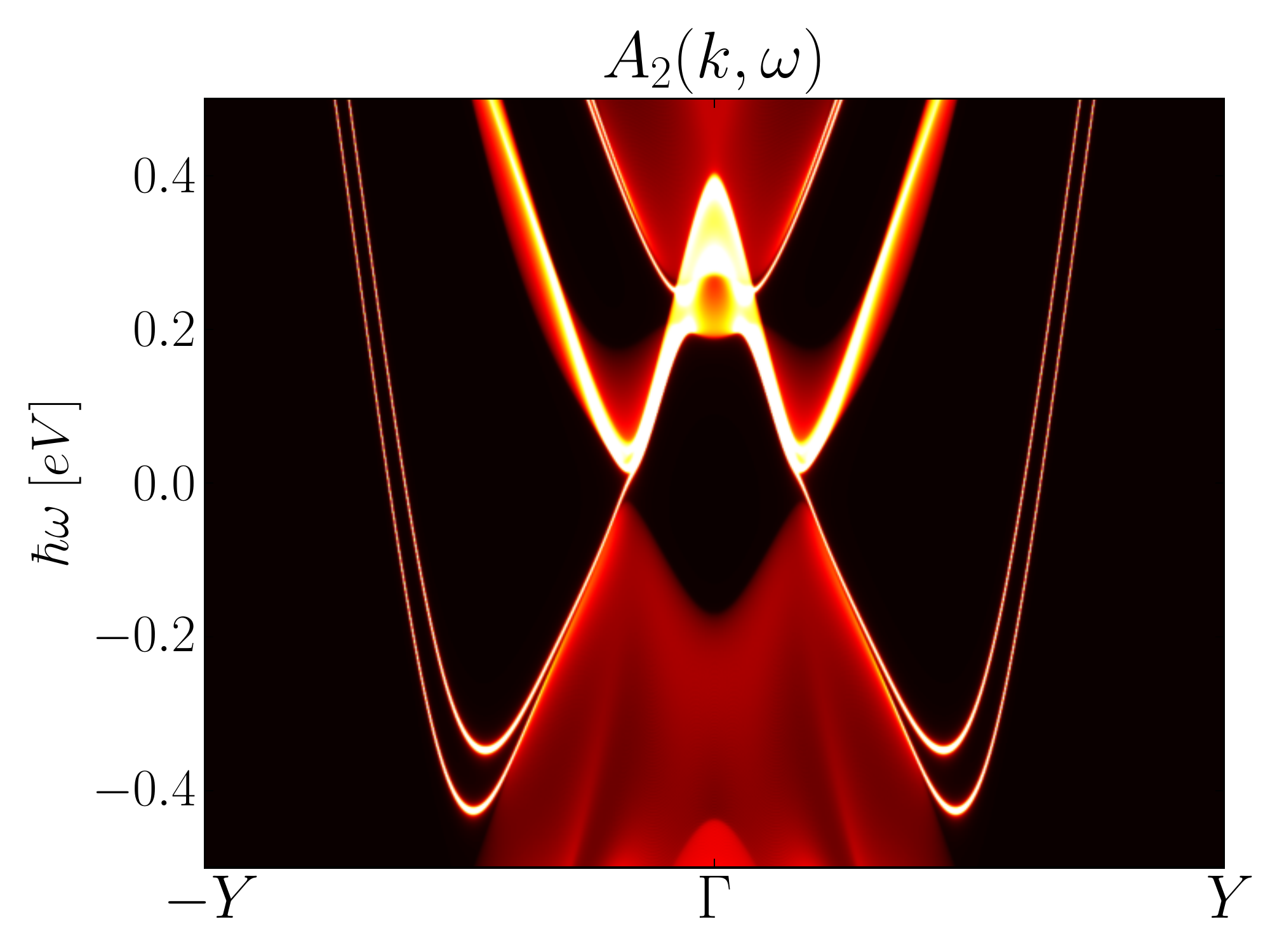}
   \includegraphics[width=5.8 cm]{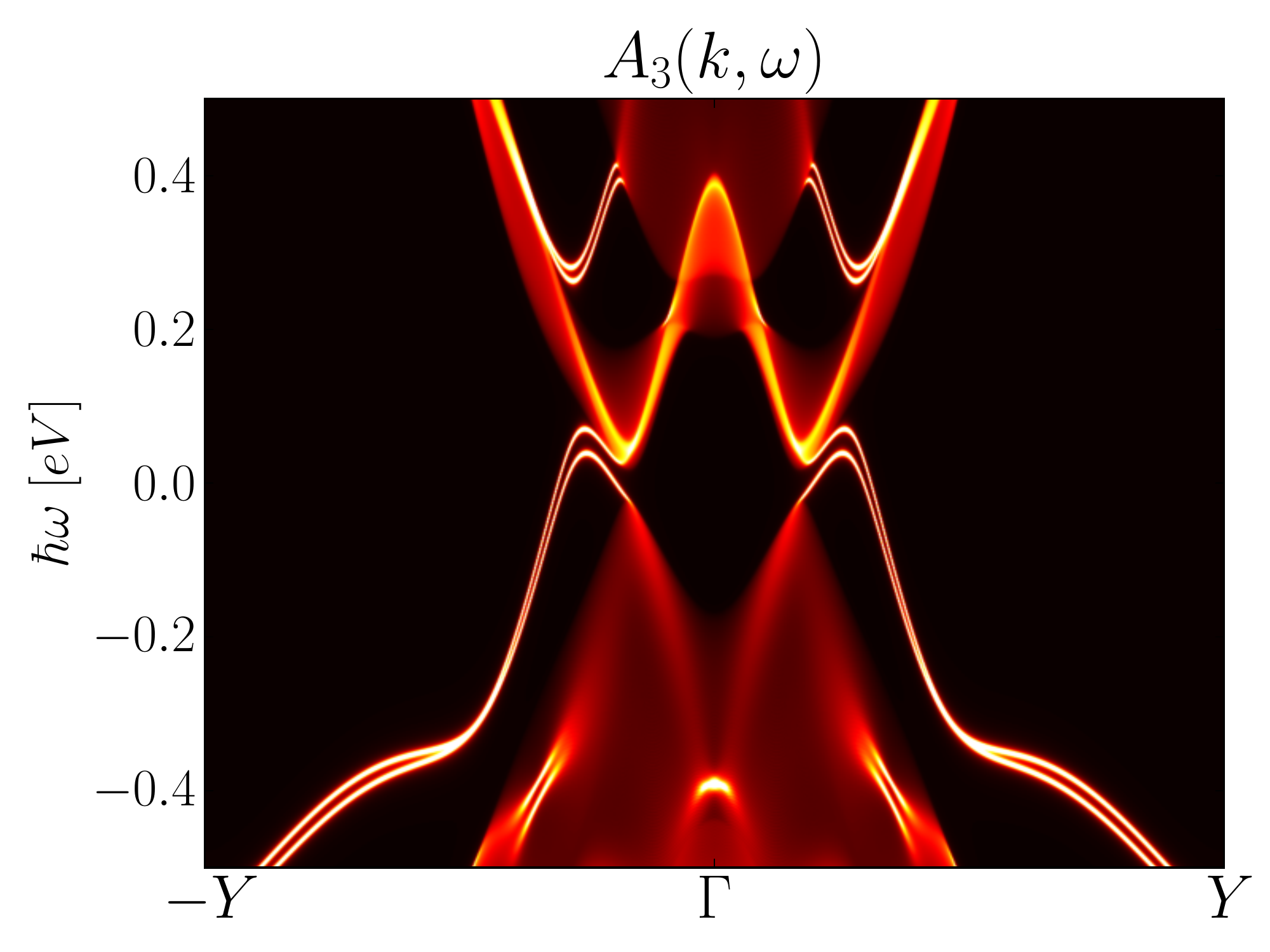}
\caption{(Color online) Spectral functions of 1T'-MoS$_2$ projected onto a single edges. The three different figures corresponds to the three edge terminations indicated in Fig. \ref{fig:bands_1T}}
\label{fig:bare_spectral}
\end{figure*}
\subsection{Pristine 1T'-MoS$_2$}
The 1T' structure of MoS$_2$ has been shown to comprise a metastable alternative to the more stable 2H-MoS$_2$ structure found in nature.\cite{Voiry2013, Lin2014, Acerce2015} The 1T structure of TMDs is constructed by forming a 2D hexagonal array of transition metal atoms and placing a hexagonal layer of chalcogen atoms in one hollow site on the top face and another hexagonal layer in another hollow site at the bottom face. For MoS$_2$ this structure is unstable and the 1T structure distorts to the 1T' structure where chains of transition metal atoms are formed in the $Y$ direction. Both the 1T and 1T' structures are inversion symmetric and whereas the 1T structure is metallic the 1T' structure is a quantum spin Hall insulator with a Kohn-Sham gap of $\sim50$ meV.\cite{Qian2014} The band structure and geometry of 1T'-MoS$_2$ is shown in Fig. \ref{fig:bands_1T}. 

Using Eq. \eqref{eq:parity}, it is straightforward to verify that the 1T'-MoS$_2$ structure has a non-trivial $\mathbb{Z}_2$ index. Since we will be calculating topological boundary states in heterostructure configurations below, it is instructive to calculate the spectrum of a few bare 1T'-MoS$_2$ edges for later reference. We thus consider the three edges obtained by the cuts indicated in Fig. \ref{fig:bands_1T}. For the present purpose we have not relaxed either the geometry or density at the edge, but simply take the bulk Hamiltonian in a local basis and remove all hopping matrix elements crossing the edges. The spectral functions at an edge can then be obtained by iterating the retarded Greens function including all hopping matrix elements within a range of two lattice vectors in the direction perpendicular to the edge.\cite{Sancho2000a} The results for the three edges are displayed in Fig. \ref{fig:bare_spectral}, which clearly shows two distinct edge states in half the edge Brillouin zone. Due to time-reversal symmetry, the pair becomes degenerate at the boundary of the Brillouin zone, but splits up and connects with the conduction and valence bands near $\Gamma$ - thus reflecting the non-trivial bulk band topology.

The strongly dispersive bands in edges 2 and 3 breaks the ideal picture of a single conducting channel and are likely to complicate the interpretation of edge states in an experimental setting where edges are typically disordered and contain contributions from several different terminations. In particular, edges 2 and 3 display three Fermi level crossings in half the Brillouin zone, however, local surface reconstruction can easily modify the dispersion and the edge can acquire any odd number of conducting channels. For the purpose of studying the 1D metallic states it is thus be desirable to stabilize the topological edge bands such that only a single band contributes to the conductivity in half the Brillouin zone. We will demonstrate below that one way of accomplishing this is to replace the edge states of QSHIs by boundary states in topological heterostructures.

\begin{table}[b]
\centering
\label{tab:atoms}
\begin{tabular}{c|c|c|c|c|c}
  & O & 2O & 2F & F$_2$ & Cl$_2$\\
 \hhline{=|=|=|=|=|=}
$E_B$ & 0.74 & 0.45 & 1.04 & 0.54 & 0.18\\
 \hline
$\Delta$ & 0.26 & 0.002 & M & M & M
\end{tabular}
\caption{Adsorption energies $E_B$ and band gaps $\Delta$ of various adsorbates on 1T'-MoS$_2$. The metallic structures are tagged with an M instead of a band gap.}

\end{table}\subsection{1T'-MoS$_2$ with adsorbates}
The small band gap of 1T'-MoS$_2$ makes it very easy to destroy the non-trivial topological phase by external perturbations. For the purpose of designing topological heterostructures like the one shown in Fig. \ref{fig:drawing}, the main challenge is therefore to prevent the perturbed material from becoming metallic. A particularly simple way of changing the electronic structure in 2D materials is by means of adsorbates. The stability of 2D materials, such as 1T'-MoS$_2$, is a consequence of the faces being rather chemically inert and they are not expected to bind adsorbates strongly.\cite{Pandey2015} Indeed, first principles calculations show that the faces of MoS$_2$ cannot bind molecules such as N$_2$, CO and CO$_2$, whereas the halogens Cl$_2$, F$_2$ and O$_2$ bind very weakly. However, most of these molecules may undergo dissociative adsorption at finite temperatures and we find that most single atoms can be bound quite strongly. In Tab. \ref{tab:atoms} we display the adsorption energies per atom of O, F and Cl relative to the pristine slab and half a dimer molecule. For Cl$_2$ and F$_2$ the binding energy is for the entire molecule. In the case of F and Cl we only consider adsorption of atoms in the unit cell since adsorption of a single atom naturally leads to a metallic state due to the odd number of electrons in these atoms. We also state the band gap in cases where the adsorbate structures is a normal insulator (None of these are quantum spin Hall insulators). In all cases we have found the minimum energy adsorbate site by relaxing the adsorbate structure from several different initial configurations. We also considered adsorption of N and H atoms (two atoms per unit cell), but did not find any configurations that are stable with respect to the dissociated molecules. In general atoms tend to adsorb on S top sites, with the S atoms residing between Mo chains being the most reactive.

Of the different adsorbates considered we found that the strongest binding energy (0.74 eV per atom) is obtained with a single O atom per unit cell. This structure has a band gap of 0.26 eV and is a normal insulator. Since the system does not have an inversion center Eq. \eqref{eq:parity} cannot be used to determine the topological index. However, no band crossings are observed while the spin-orbit coupling is adiabatically turned off, which implies that the system is a trivial insulator. We will investigate this adsorbate structure further in the following.

\begin{figure}[tb]
   \includegraphics[width=4.2 cm]{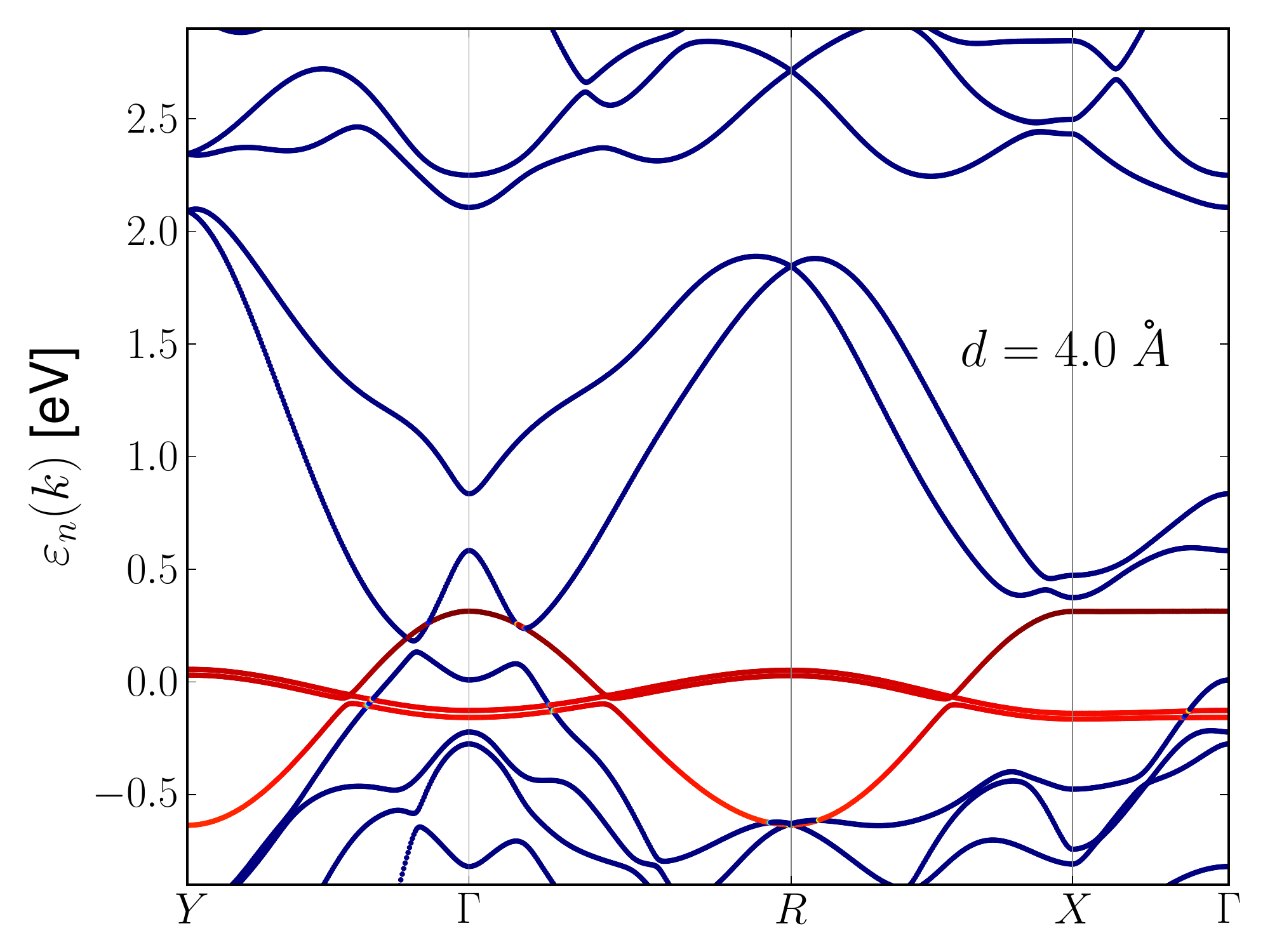}
   \includegraphics[width=4.2 cm]{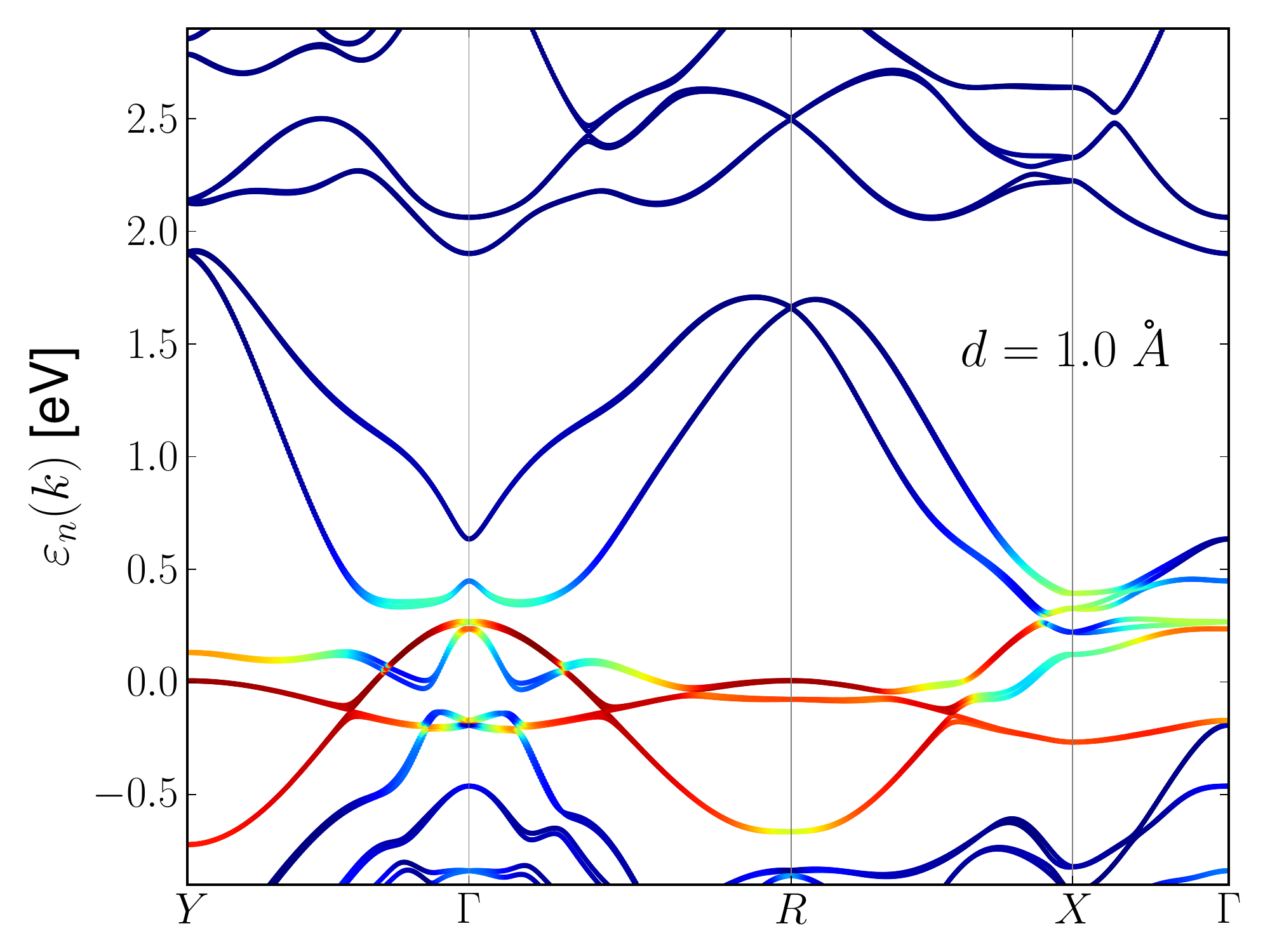}
   \includegraphics[width=4.2 cm]{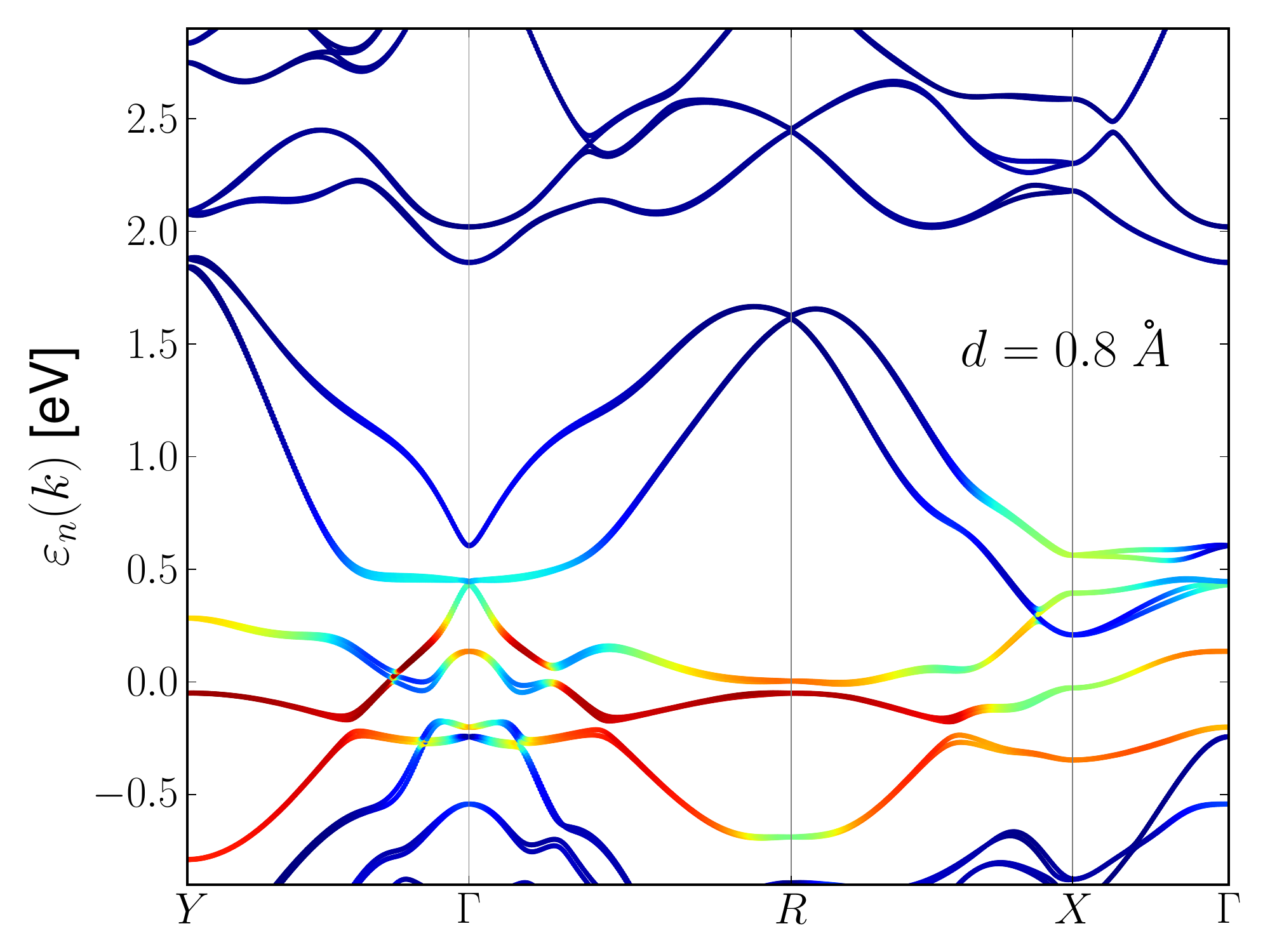}
   \includegraphics[width=4.2 cm]{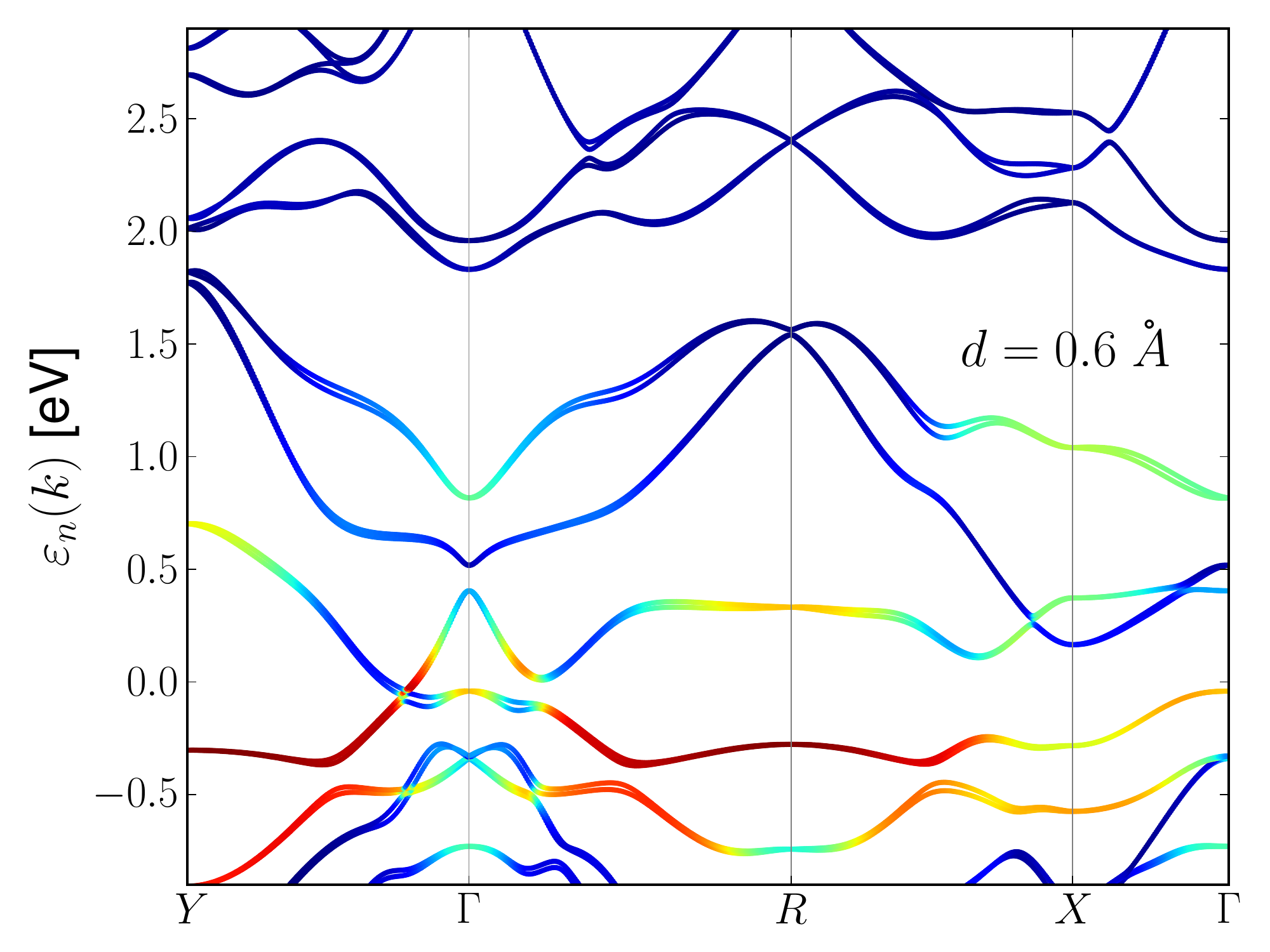}
   \includegraphics[width=4.2 cm]{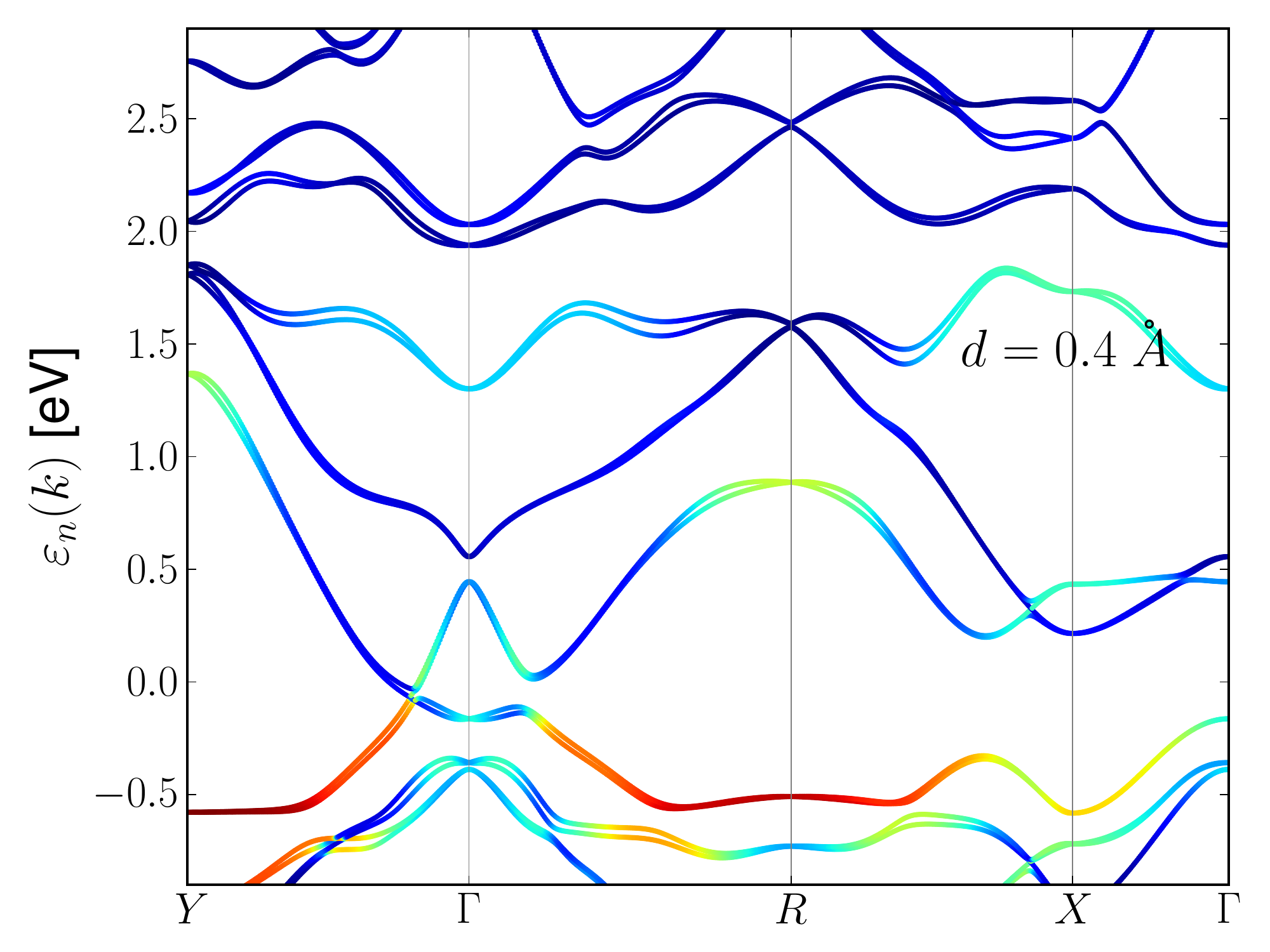}
   \includegraphics[width=4.2 cm]{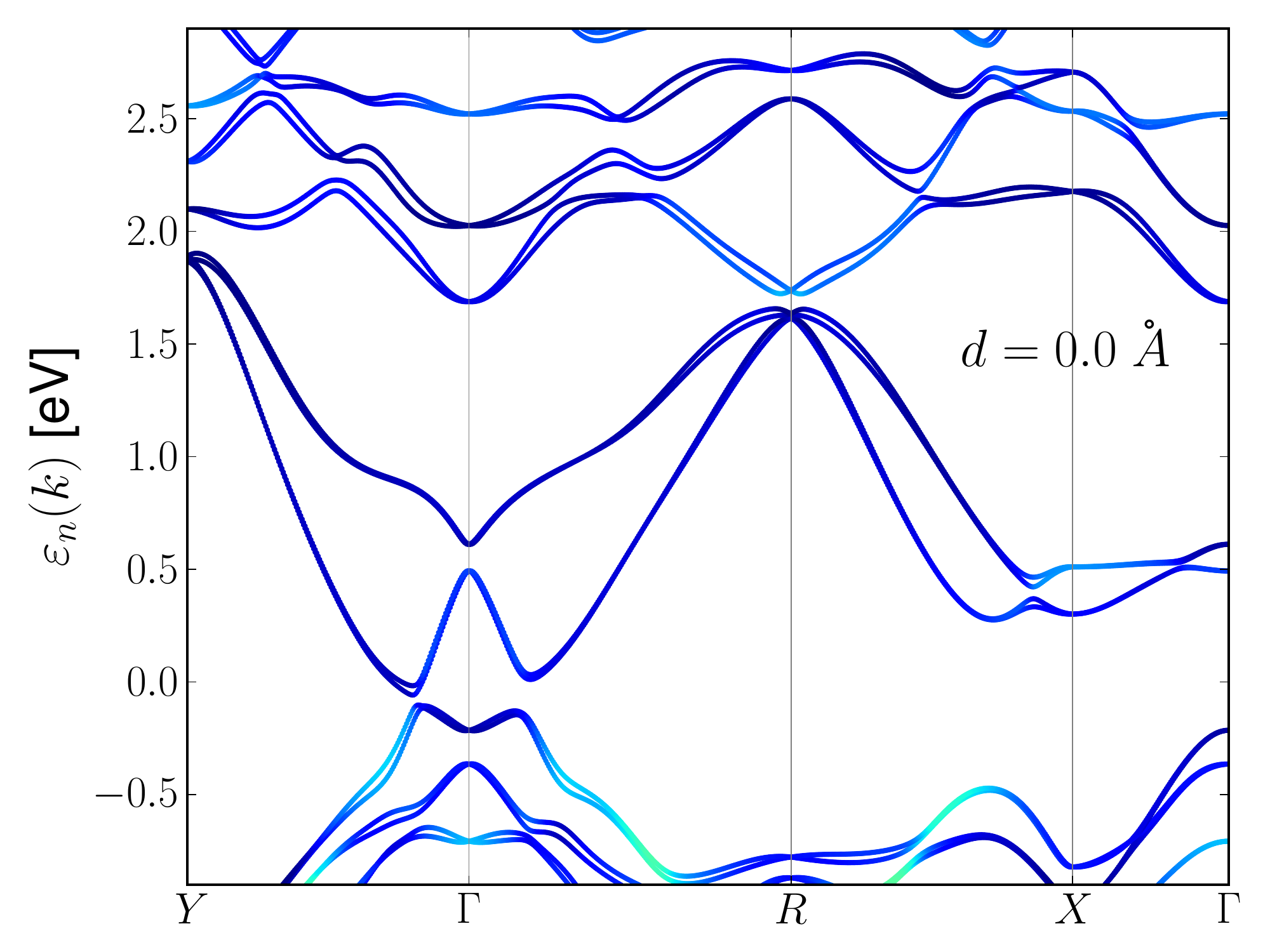}
\caption{(Color online) Band structure at different snapshots along the configuration path where a single O atom adsorbs on the face of 1T'-MoS$_2$. The red color indicates oxygen character of the bands and the blue color indicates 1T'-MoS$_2$ character.}
\label{fig:dissociation}
\end{figure}
\subsubsection{Transition from non-trivial to trivial topology upon adsorption}
It is instructive to follow the transition from a topological insulator to a trivial insulator as an oxygen atoms is approaching the layer. The isolated oxygen atom at half coverage is weakly interacting with their nearest neighbors and the oxygen array itself thus comprises a weakly dispersive metallic system. Initially, when the O atom is far away, the system can then be viewed as a metal superimposed on the 2D topological insulator. It is the six O $p$ bands that participate in hybridization near the Fermi level and four of the bands thus have to enter the valence manifold and two bands have to enter the conduction manifold if the adsorbate structure is to end up as an insulator.

In Fig. \ref{fig:dissociation}, we follow the band structure of the combined system as an O atom is moved towards the equilibrium adsorption point. Strictly speaking, the system becomes spin-polarized when the O atom is moved far from the slab and the system thus exhibits a transition from being spin-polarized to spin-paired at a point along the adsorption path. However, for simplicity we have only considered a spin-paired desorption path here. When the O atom is approaching the surface, the O $p$ bands start to hybridize weakly with the 1T'-MoS$_2$ bands.  At that point the entire system is metallic and we can no longer regard it as a metallic array of O atoms superimposed on a topological insulator. Although the insulating nature of the system is not restored until the O atom reaches its equilibrium position the figures hints at the mechanism leading to a change in topology. In the region of $d=1.0-0.6$ {\AA} an additional conduction band emerges which has 1T'-MoS$_2$ character at $\Gamma$. Since this originated from the original frontier valence orbitals, it is clear that the band inversion that led to a non-trivial topology in the pristine system has been transferred to the conduction band. We note the strong similarity between the band structure of pristine 1T'-MoS$_2$ shown in Fig. \ref{fig:bands_1T} and the trivial insulating band structure obtained with adsorbed oxygen ($d=0$).

\begin{figure}[tb]
   \includegraphics[width=6.0 cm]{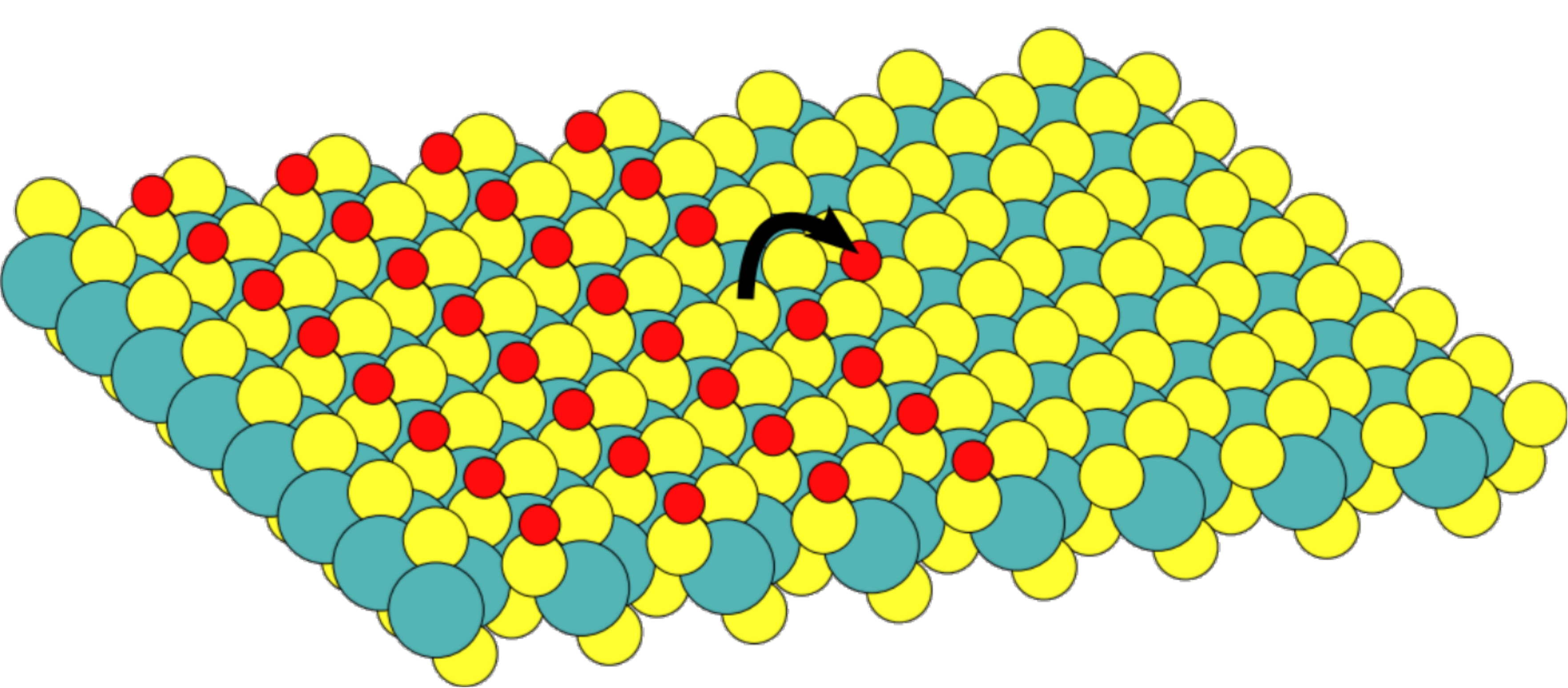}
\caption{(Color online) Oxygen diffusion used for the nudged elastic band calculation. The transition barrier for the hopping is 2 eV.}
\label{fig:neb}
\end{figure}
\begin{figure*}[tb]
   \includegraphics[width=5.8 cm]{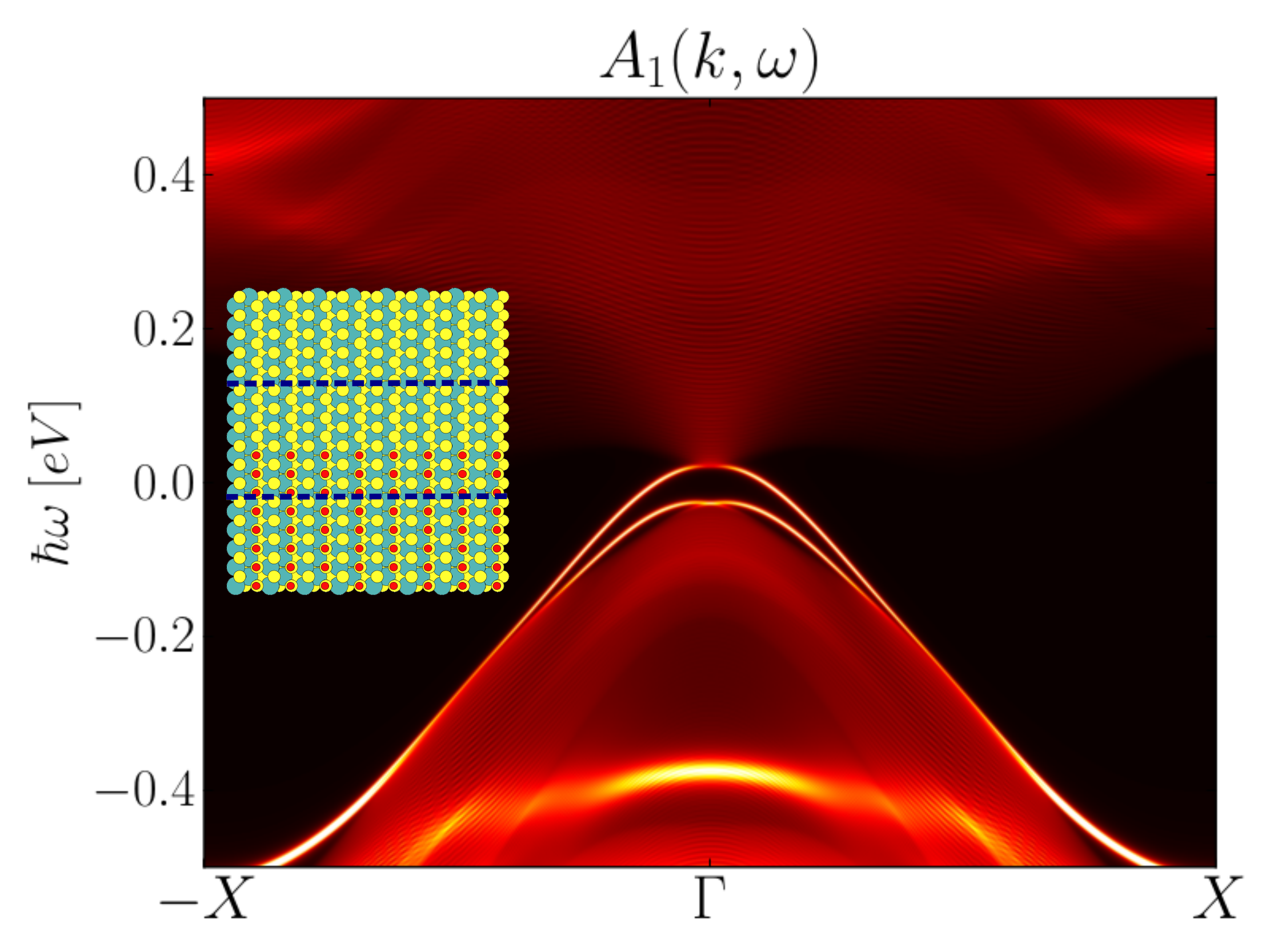}
   \includegraphics[width=5.8 cm]{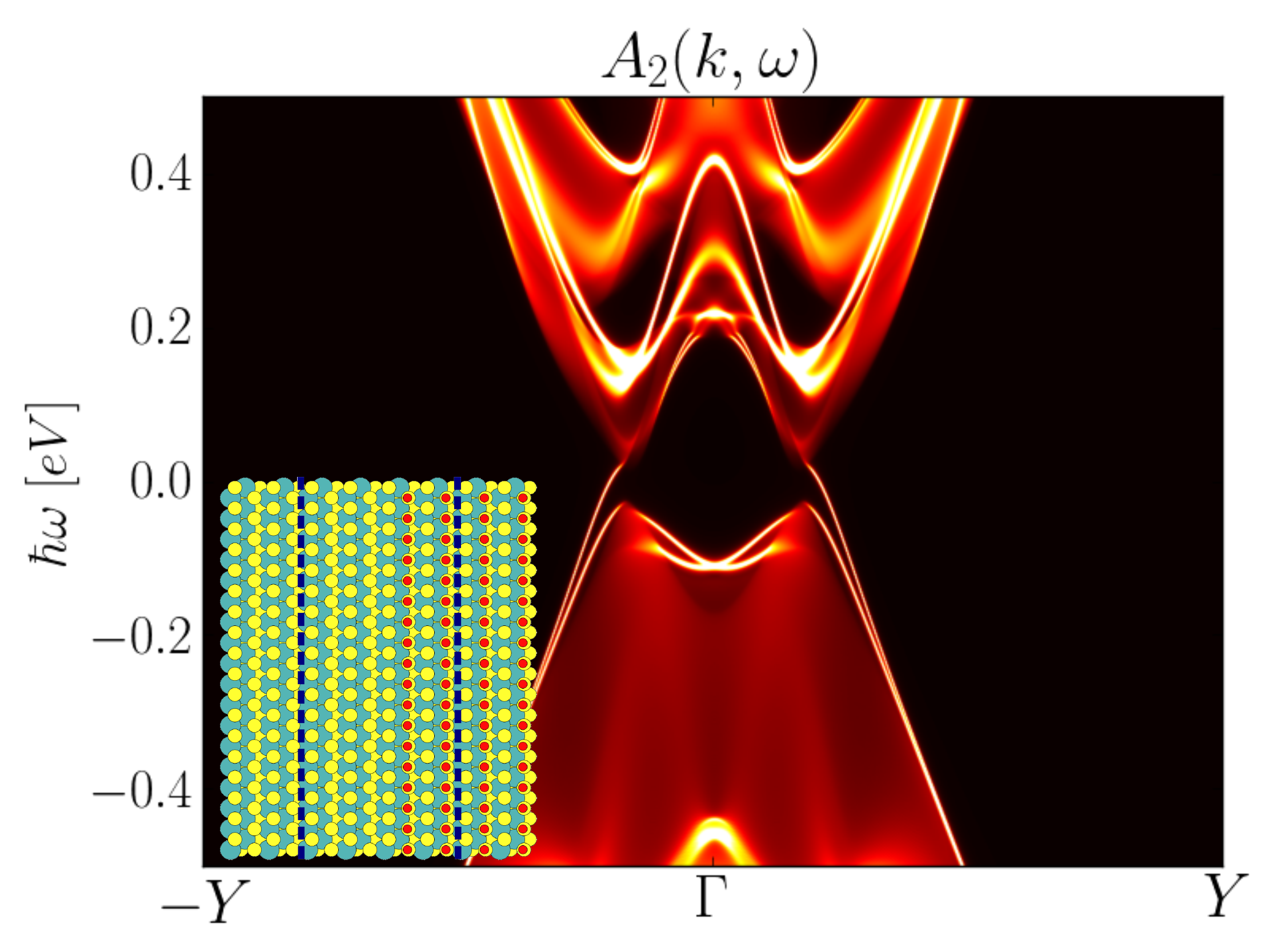}
   \includegraphics[width=5.8 cm]{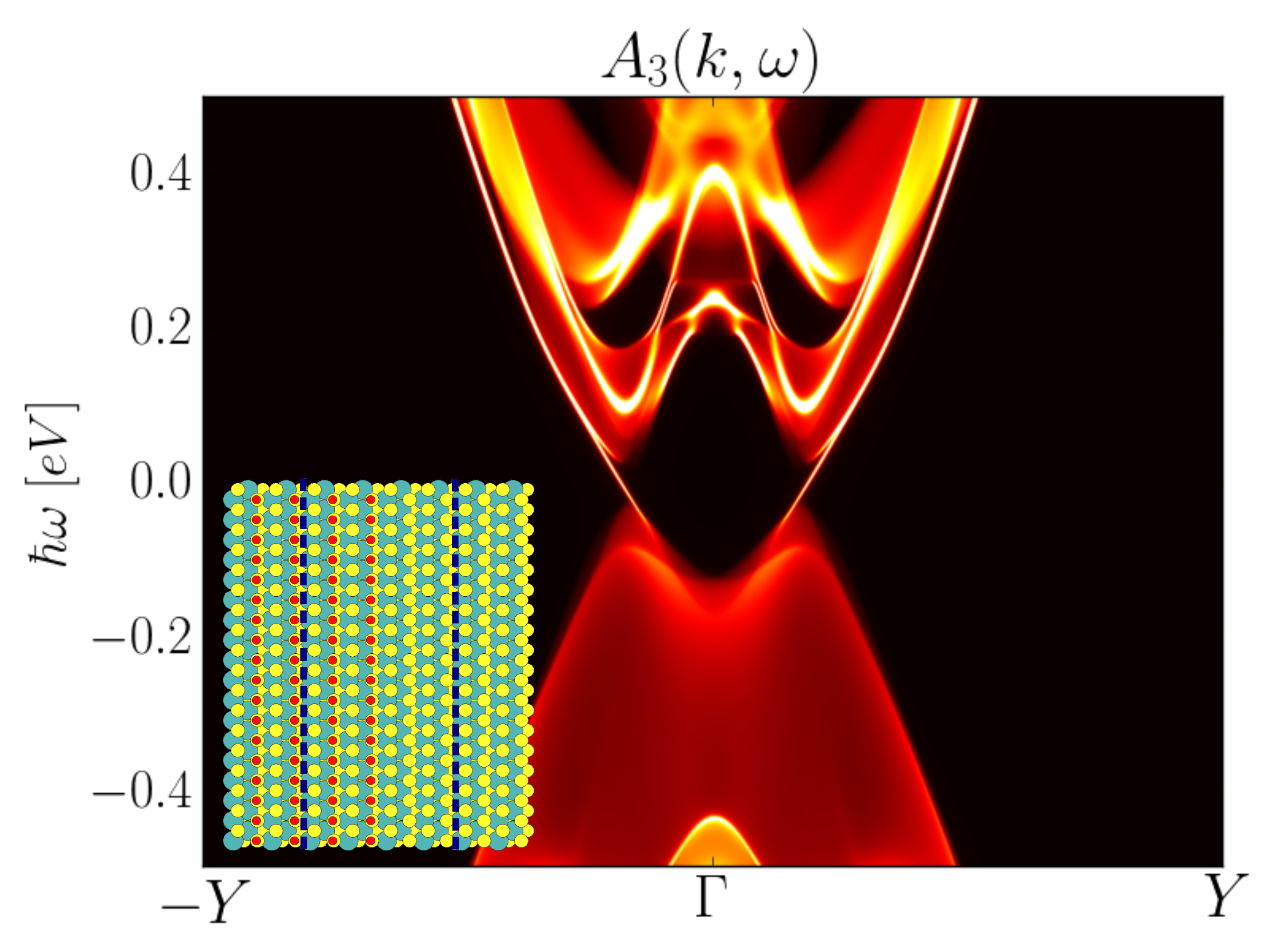}
\caption{(Color online) Left: Spectral function of 1T'-MoS$_2$ heterostructures with a semi-infinite adsorbate layer of oxygen. The three different figures have different terminations of the adsorbate layer and correspond roughly to the three figures in Fig. \ref{fig:bare_spectral}.}
\label{fig:hetero_spectral}
\end{figure*}
\subsubsection{Stability of heterostructure}
Although atomic oxygen binds quite strongly to the 1T'MoS$_2$ sheets, it is not a priori clear if an interface of pristine 1T'-MoS$_2$ and 1T'-MoS$_2$O will be stable, since oxygen may diffuse from the adsorbate region to the pristine region. This is illustrated in Fig. \ref{fig:neb} where a hopping event between two nearest neighbor adsorption sites is indicated. In order to calculate the barrier for such an event we invoke the climbing nudged elastic band method\cite{Henkelman2000} to obtain the transition state. We find a diffusion barrier of 2.2 eV indicating that the interface is highly stable.

\subsubsection{Boundary states of heterostructures}
In Fig. \ref{fig:hetero_spectral} we show the spectral functions of three different boundary regions of MoS$_2$ heterostructures with semi-infinite adsorbate layers of O. In each case the boundary region is indicated by the insert and the three boundaries correspond roughly to the three cuts shown in Fig. \ref{fig:bare_spectral}. In order to obtain a fully relaxed boundary we have considered systems that are comprised of 8 repetitions of the unit cell in the direction orthogonal to the boundary and half of the unit cells contain adsorbed oxygen. We then performed a full relaxation of the interface region consisting of two pristine unit cells and two unit cells with adsorbates. Finally, the hopping parameters connecting the boundary with the bulk were obtained from this structure and the bare bulk hopping parameters were obtained from calculations of pristine 1T'-MoS$_2$ with and without adsorbates. We refer to appendix \ref{app:greens} for details. 

All the structures exhibit very well-behaved topological surface states with a single Fermi level crossing in the half the Brillouin zone. In particular, the strongly dispersive edge states originating from cuts 2 and 3 in Fig. \ref{fig:bare_spectral} have disappeared and are replaced by minimalistic connections between valence and conduction band continua. We thus expect such topological boundary states to be very stable towards boundary reconstruction and disorder, which strongly facilitates experimental control of these 1D metallic states. This is in sharp contrast to the case of topological edge states, which are highly sensitive to the details of edge termination.

\section{Conclusion}
In conclusion, we have demonstrated that first principles simulations provide an easy means to design topological heterostructures with one-dimensional boundary states. While it is straightforward to study 1T'-MoS$_2$ heterostructures by means of computer simulations, it might be a completely different matter to do so experimentally. Although 1T'-MoS$_2$ has been isolated and characterized experimentally\cite{Voiry2013, Lin2014, Acerce2015} the material is metastable and will eventually decay to the 2H structure. Moreover, it is by no means clear that local adsorbate regions can be obtained by standard techniques and the construction of the boundary regions considered in the present work could be non-trivial. Nevertheless, even if it turns out to be impractical to work with the 1T'-MoS$_2$-O system experimentally, there are several other QSHI-adsorbate structures that could be proposed and it is highly likely that new and more stable QSHIs will be discovered in the near future. In addition, the example of 1T'-MoS$_2$ explicitly demonstrates two aspects of QSHI heterostructures that we expect to be rather generic. First, the non-trivial topology of QSHIs is easily changed upon adsorption of atoms and molecules, which suggests that topological heterostructures can easily be constructed using a pristine QSHI with local areas of adsorbates. Second, the boundary states in these heterostructures do not show the strong dispersion and multiple Fermi level crossings characteristic of QSHI edge states and thus provide a much better platform for studying topologically protected conductivity in 1D.

\section{Acknowledgement}
The Center for Nanostructured Graphene (CNG) is sponsored by the Danish National Research Foundation, Project DNRF58.

\appendix

\section{Implementation of spin-obit coupling in GPAW}\label{app:so}
In this appendix, we will provide details on the spin-orbit implementation in the electronic structure code GPAW. The implementation is based on a non-selfconsistent diagonalization of the Kohn-Sham Hamiltonian including the spin-orbit interaction. We thus consider the full Hamiltonian in a basis of scalar-relativistic Kohn-Sham eigenstates:
\begin{align}\label{eq:H_so}
 H_{n_1n_2\sigma_1\sigma_2}=\varepsilon_{n_1\sigma_1}\delta_{n_1n_2}\delta_{\sigma_1\sigma_2}+\langle\psi_{n_1\sigma_1}|\hat H_{SO}|\psi_{n_2\sigma_2}\rangle
\end{align}
where the spin-orbit Hamiltonian is given by
\begin{align}
 H_{SO}(\mathbf{r})=\frac{\hbar\boldsymbol{\sigma}\cdot\mathbf{p}\times\boldsymbol{\nabla}v_{KS}(\mathbf{r})}{4m^2c^2},	
\end{align}
$v_{KS}$ is the spin-independent part of the Kohn-Sham potential, and $\varepsilon_{n\sigma}$ are the self-consistent eigenvalues of the scalar-relativistic Kohn-Sham Hamiltonian. Due to the derivative of the Kohn-Sham potential, the spin-orbit correction is completely dominated by the regions close to the nuclei in atomic systems. In the projector-augmented wave (PAW) formalism, we can thus restrict the evaluation of the correction to regions inside the PAW spheres.\cite{blochl, Enkovaara2010a} In these regions the all-electron orbitals can be expanded as
\begin{align}
 |\psi_{n\sigma}\rangle=\sum_i\langle\tilde p^a_{i\sigma}|\tilde\psi_{n\sigma}\rangle|\phi_{i\sigma}^a\rangle,
\end{align}
where $|\phi_{i\sigma}^a\rangle$ are the all-electron partial waves, $|\tilde p_{i\sigma}^a\rangle$ are their dual projectors and $|\tilde\psi_{n\sigma}^a\rangle$ are the smooth pseudo-wavefunctions. Here $a$ is an index denoting a particular augmentation sphere. We can thus write
\begin{align}
&\langle\psi_{n_1\sigma_1}|\hat H_{SO}|\psi_{n_2\sigma_2}\rangle =\notag\\ 
&\sum_{ai_1i_2}\langle\tilde\psi_{n_1\sigma_1}|\tilde p^a_{i_1\sigma_1}\rangle\langle\phi_{i_1\sigma_1}^a|\hat H_{SO}|\phi_{i_2\sigma_2}^a\rangle\langle\tilde p^a_{i_2\sigma_2}|\tilde\psi_{n_2\sigma_2}\rangle,
\end{align}
where we neglected cross contributions from different augmentation spheres. The projector overlaps $\langle\tilde p^a_{i\sigma}|\tilde\psi_{n\sigma}\rangle$ are calculated during any standard Kohn-Sham calculation and are readily available. We are therefore left with a calculation of the partial wave contributions $\langle\phi_{i\sigma}^a|\hat H_{SO}|\phi_{i\sigma}^a\rangle$.

To proceed, we note that the dominant contribution to the potential entering the spin-orbit correction originates from the bare nuclei and frozen electronic core, which gives rise to spherically symmetric potentials. We thus assume a spherically symmetric form of the  spin-orbit Hamiltonian. Decomposing the partial wave as a spherical harmonic $|Y^a_i\rangle$, a radial function $|f^a_i\rangle$ and a spinor $|\sigma\rangle$ we obtain
\begin{align}
&\langle\phi_{i_1\sigma_1}^a|\hat H_{SO}|\phi_{i_2\sigma_2}^a\rangle=\\
&-\frac{1}{2m^2c^2}\langle Y_{i_1}^a\sigma_1|\mathbf{\hat S}\cdot\mathbf{\hat L}|Y_{i_2}^a\sigma_2\rangle\langle f_{i_1}^a|\frac{1}{r}\frac{d\hat v_{KS}}{dr}|f_{i_2}^a\rangle\notag.
\end{align}
The first matrix element is straightforward to evaluate analytically, since the angular momentum operator $\mathbf{\hat L}$ is easily expressed in a basis of spherical harmonics and the spin operator $\mathbf{\hat S}$ is easily applied to the spinors once a quantization axis for the spins is supplied. The second matrix element is evaluated numerically on a radial non-uniform grid with the nucleus at the origin. We note that in addition to the spin-orbit eigenvalues, the diagonalization of \eqref{eq:H_so} will yield the spinorial eigenstates in a basis of scalar-relativistic states, from which the spinorial wavefunctions can be constructed in real space.

\begin{figure}[tb]
   \includegraphics[width=7.5 cm]{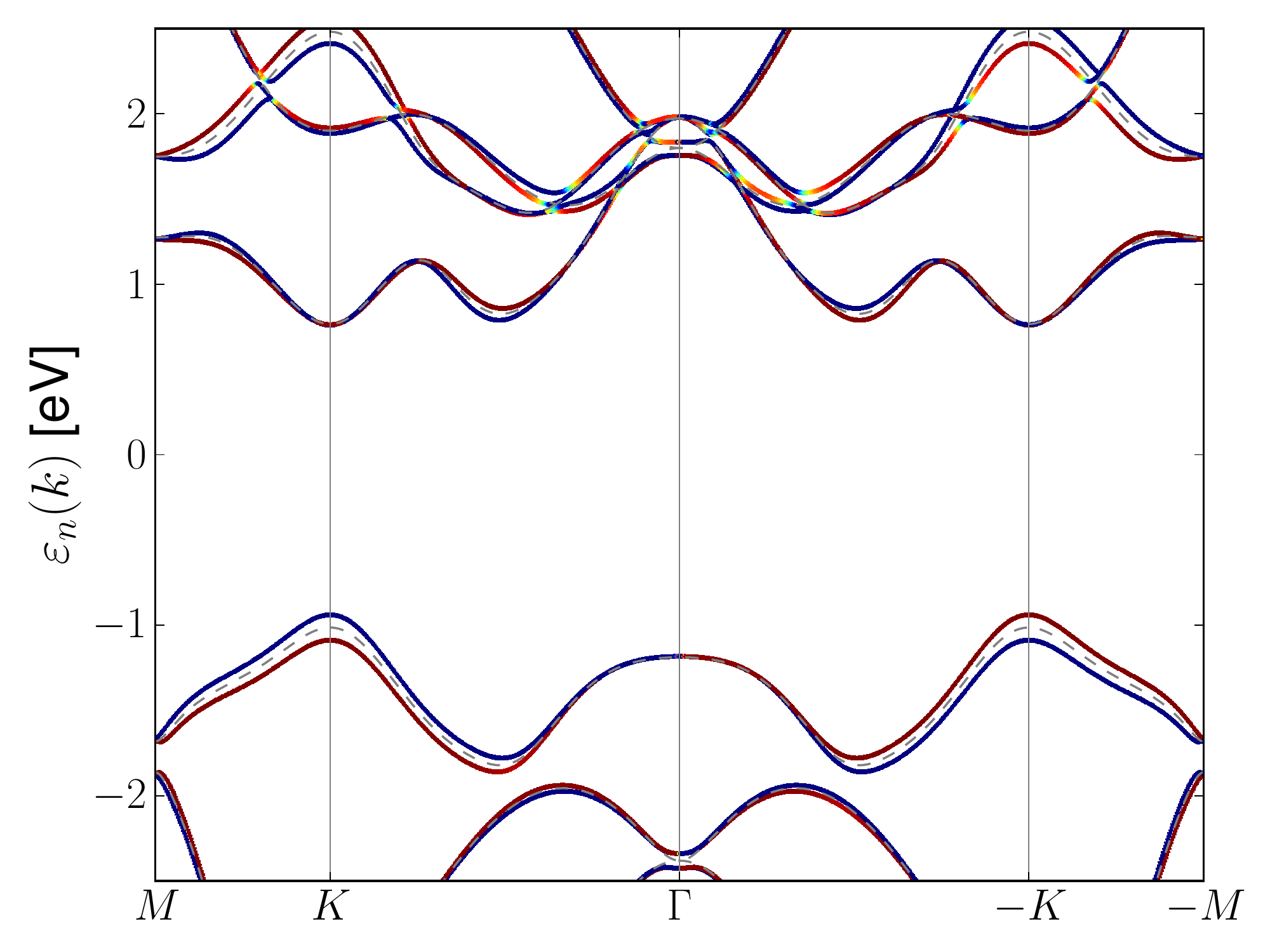}
\caption{(Color online) Band structure of 2H-MoS$_2$. The colors indicate the spin character $S_{nk}=\langle nk|\sigma_z|nk\rangle$, with blue being spin down ($S_{nk}$=-1) and red being spin up ($S_{nk}$=1), The bands without spin-orbit coupling are indicated by dashed grey lines.}
\label{fig:bands_2H}
\end{figure}

As a first test of the implementation we calculate the bandstructure of 2H-MoS$_2$ which is a trivial insulator. The 2H structure does not have inversion symmetry and as a consequence the Kramers degeneracy at individual $k$-points is lifted. The band structure is shown in Fig. \ref{fig:bands_2H} and we observe a 0.1497 eV splitting of the valence bands at $K$. This in very good agreement with previous calculations\cite{Zhu2011, Cheiwchanchamnangij2012, Olsen2015} and experiments.\cite{Mak2010} Replacing the full Kohn-Sham potential with the bare core yield a spin-orbit splitting of 0.1940 meV at $K$. If we add the spherically symmetric electronic core density we obtain 0.1490 eV, which is very close to the value obtained from the full Kohn-Sham potential. The spin-orbit coupling is thus dominated by the contribution from the bare nucleus and nearly completely captured if we include the core electrons. This indicates that the evaluation the spin-orbit coupling inside the augmentation spheres and the restriction to the spherical components of the valence density comprises a highly accurate approach.

\begin{figure}[tb]
   \includegraphics[width=7.5 cm]{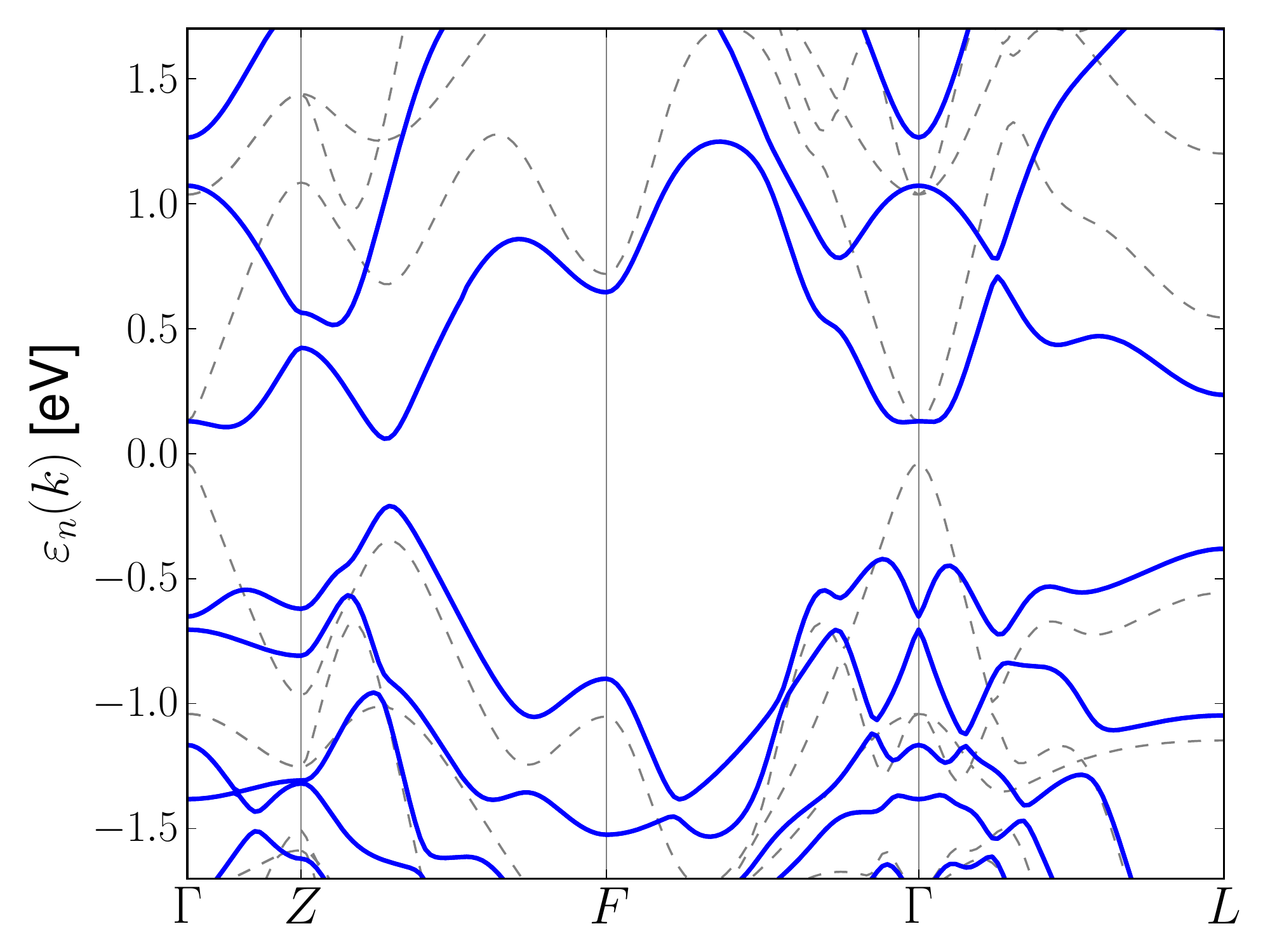}
\caption{(Color online) Band structure of Bi$_2$Se$_3$ with (blue lines) and without (dashed grey lines) spin-orbit coupling.}
\label{fig:bands_Bi2Se3}
\end{figure}

As a second example, we consider the inversion symmetric strong topological insulator Bi$_2$Se$_3$. The band structure is shown in Fig. \ref{fig:bands_Bi2Se3} with and without spin-orbit coupling and is in very good agreement with previous calculations.\cite{Zhang2009} In order to demonstrate that we obtain the correct non-trivial band topology, we calculate the $\mathbb{Z}_2$ index $\nu$ using the expression for parity invariant systems\cite{Fu2007}
\begin{align}\label{eq:parity}
(-1)^{\nu}=\prod_a\prod_m\xi_m(\Lambda_a),
\end{align}
where $m$ runs over occupied Kramers pairs at each of the time-reversal invariant momenta $\Lambda_a$. We find that the product of parity eigenvalues changes from $-1$ to $1$ at $\Gamma$ upon inclusion of spin-orbit coupling and the $\mathbb{Z}_2$ index changes from $\nu=0$ to $\nu=1$ accordingly. One can follow the topological transition by replacing the spin-orbit interaction $H_{SO}$ by $\lambda H_{SO}$ and adiabatically tuning $\lambda$ from 0 to 1. We observe a transition through a metallic state at $\lambda=0.35$, which marks the transition between two insulating states of different topology.

\section{GPAW-Wannier90 interface}\label{app:wannier}
The construction of maximally localized Wannier functions makes use of the Bloch function overlaps\cite{Marzari2012}
\begin{align}
 M_{mn}^{\mathbf{k},\mathbf{b}} = \langle u_{m\mathbf{k}}|u_{n\mathbf{k+b}}\rangle=\langle\psi_{m\mathbf{k}}|e^{-i\mathbf{b}\cdot\mathbf{\hat r}}|\psi_{n\mathbf{k+b}}\rangle,
\end{align}
where $\mathbf{b}$ are a set of vectors that connects $\mathbf{k}$ to the nearest neighbor $k$-points. Within the PAW formalism these can be evaluated as 
\begin{align}\label{eq:M_mn}
 M_{mn}^{\mathbf{k},\mathbf{b}} = &\langle\tilde\psi_{m\mathbf{k}}|e^{-i\mathbf{b}\cdot\mathbf{\hat r}}|\tilde\psi_{n\mathbf{k+b}}\rangle\\
 +&\sum_{a,i,j}e^{-i\mathbf{b}\cdot\mathbf{r}_a}\langle\tilde\psi_{m\mathbf{k}}|\tilde p_i^a\rangle\Big(\langle\phi_i^a|\phi_j^a\rangle-\langle\tilde\phi_i^a|\tilde\phi_j^a\rangle\Big)\langle\tilde p_j^a|\tilde\psi_{n\mathbf{k+b}}\rangle\notag,
\end{align}
where we have neglected overlap contributions from neighboring PAW spheres and approximated
\begin{align}
\langle\phi_i^a|e^{-i\mathbf{b}\cdot\mathbf{\hat r}}|\phi_j^a\rangle\approx\langle\phi_i^a|\phi_j^a\rangle e^{-i\mathbf{b}\cdot\mathbf{r}_a}.
\end{align}
The first term in Eq. \eqref{eq:M_mn} is smooth and can be evaluated on a real space grid, whereas all the factors entering the second term are used during an ordinary DFT calculations in the PAW formalism and can be extracted without additional computational cost.

The construction of Wannier functions also requires an initial projection onto a set of localized states. In the PAW formalism it is natural to use the set of partial waves $\phi^a_ i$ associated with each atom $a$. We then simple use
\begin{align}
A_{ain}^{\mathbf{k}}=\langle\phi^a_i|\psi_{n\mathbf{k}}\rangle\approx\langle\tilde p^a_i|\tilde\psi_{n\mathbf{k}}\rangle,
\end{align}
which is calculated during any Kohn-Sham iteration and can be obtained without additional computational cost. By forming the relevant linear combinations of the projector overlaps, it becomes possible to project onto sp$^3$ sp$^3$d$^2$ orbitals and so forth.

\section{Iterative scheme for the boundary spectral function}\label{app:greens}
In Ref. \onlinecite{Sancho2000a}, the authors presented a rapidly converging scheme for obtaining surface Greens functions in a localized basis. Here we present a simple generalization of the approach for a boundary region. The equations below are straightforward to derive following the steps in Ref. \onlinecite{Sancho2000a}.

\begin{figure}[tb]
   \includegraphics[width=7.5 cm]{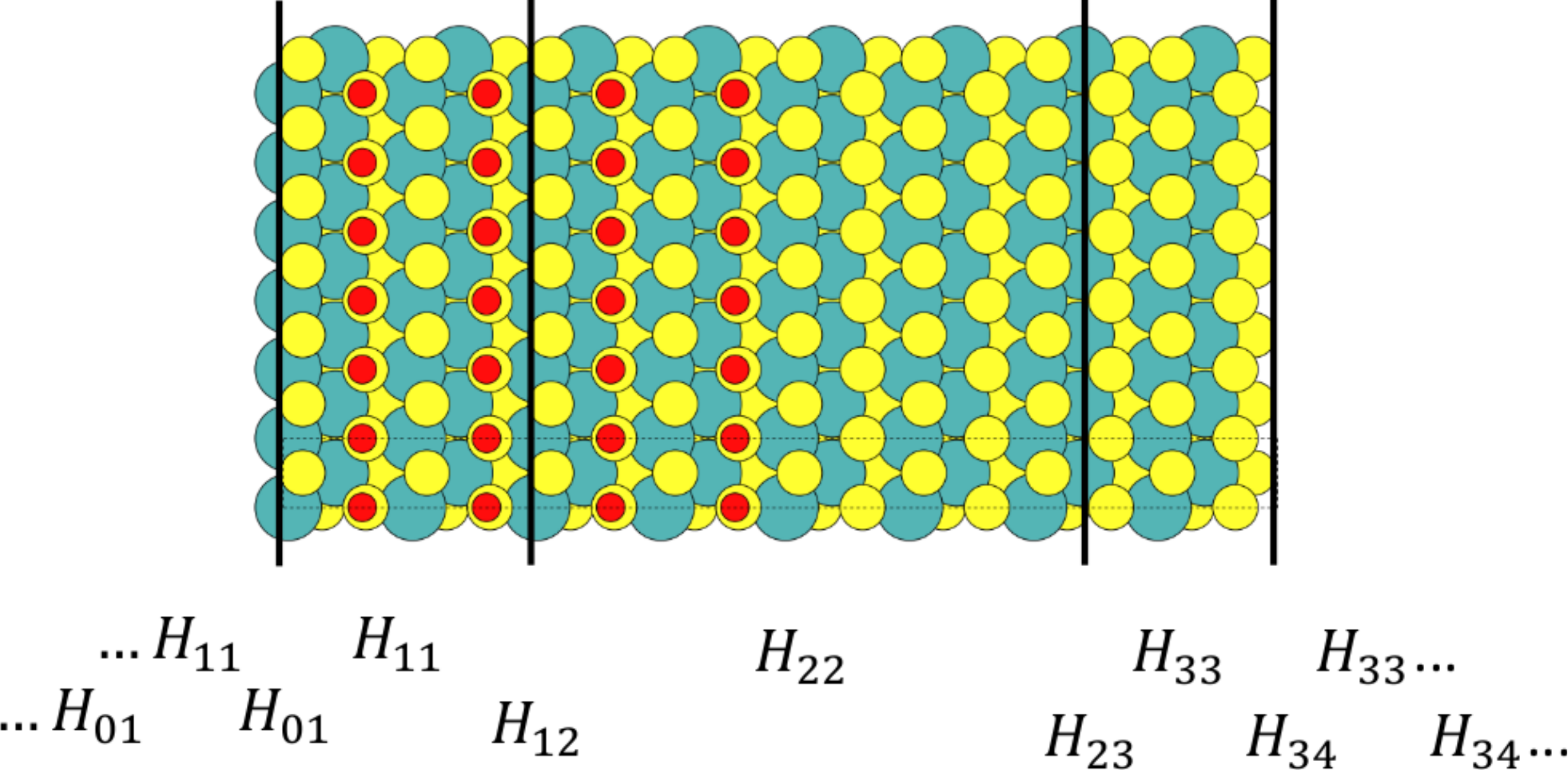}
\caption{(Color online) Division of regions used for the iterative Greens function calculation of the interface spectral function.}
\label{fig:greens}
\end{figure}
The spectral function of the boundary is obtained as 
\begin{equation}\label{eq:spectral}
 A_B(\omega, \mathbf{k}_\parallel)=\text{Tr}_\text{B}\Big[\text{Im}G^R(\omega,\mathbf{k}_\parallel)\Big],
\end{equation}
where $G^R$ is the retarded Greens function, $\text{Tr}_\text{B}$ denotes a trace over the boundary region and $\mathbf{k}_\parallel$ is the Bloch momentum parallel to the interface. In order to perform the trace one needs to express the Greens function in a local basis. However, the Greens function in the boundary region cannot be obtained from the boundary region alone, but can be calculated once the coupling to repetitive bulk regions are known. Specifically, an interface region may be divided into three regions (indicated in Fig. \ref{fig:greens}) described by the local Hamiltonians $H_{ab}(\mathbf{k}_\parallel)$ and iterate the following equations
\begin{align}
 \alpha_{i+1}&=\alpha_iG^{3}_i\beta_i,\\
 \tilde\alpha_{i+1}&=\tilde\beta_iG^{3}_i\tilde\alpha_i,\\
 \beta_{i+1}&=\beta_iG^{3}_i\beta_i,\\
 \tilde\beta_{i+1}&=\tilde\beta_iG^{3}_i\tilde\beta_i,\\
 \gamma_{i+1}&=\gamma_iG^{1}_i\delta_i,\\
 \tilde\gamma_{i+1}&=\tilde\delta_iG^{1}_i\tilde\gamma_i,\\
 \delta_{i+1}&=\delta_iG^{1}_i\delta_i,\\
 \tilde\delta_{i+1}&=\tilde\delta_iG^{1}_i\tilde\delta_i,\\
 \varepsilon_{i+1}^1&=\varepsilon_i^1+\delta_iG^1_i\tilde\delta_i+\tilde\delta_iG^1_i\delta_i,\\
 \varepsilon_{i+1}^2&=\varepsilon_i^2+\alpha_iG^3_i\tilde\alpha_i+\tilde\gamma_iG^1_i\gamma_i,\\
 \varepsilon_{i+1}^3&=\varepsilon_i^3+\beta_iG^3_i\tilde\beta_i+\tilde\beta_iG^3_i\beta_i
\end{align}
where
\begin{align}
 G^a_i(\omega)=(\omega-\varepsilon^a_i+i\eta)^{-1}.
\end{align}
In the present work we have taken $H_{01}$, $H_{11}$, $H_{33}$, and $H_{34}$ from bulk calculations of the pristine systems present at the two sides of the boundary, whereas $H_{12}$, $H_{22}$, and $H_{23}$ is calculated from boundary structure. The initial conditions for the equations are then
\begin{align}
 \alpha_0=H_{23},\qquad\tilde\alpha_0=H_{23}^\dag,\\
 \beta_0=H_{34},\qquad\tilde\beta_0=H_{34}^\dag,\\
 \gamma_0=H_{12},\qquad\tilde\gamma_0=H_{12}^\dag,\\
 \delta_0=H_{01},\qquad\tilde\delta_0=H_{01}^\dag,
\end{align}
and
\begin{align}
 \varepsilon_0^1=H_{11},\quad\varepsilon_0^2=H_{22},\quad\varepsilon_0^3=H_{33}.
\end{align}
The boundary Greens function then contains everything needed for the trace in Eq. \eqref{eq:spectral} and can be obtained from
\begin{align}
 G^B(\omega)=G^2_{i\rightarrow\infty}(\omega).
\end{align}


\end{document}